\documentclass[conference, 9pt]{IEEEtran}
\usepackage{spconfa4,graphicx}
\usepackage{graphicx}%
\usepackage{multirow}%
\usepackage{amsmath,amssymb,amsfonts}%
\usepackage{amsthm}%
\usepackage{mathrsfs}%
\usepackage[figuresright]{rotating}%
\usepackage{xcolor}%
\usepackage{textcomp}%
\usepackage{manyfoot}%
\usepackage{booktabs}%
\usepackage{algorithm}%
\usepackage{algorithmicx}%
\usepackage{algpseudocode}%
\usepackage{program}%
\usepackage{listings}%
\usepackage{cite}%

\begin{document}

\title{Adaptive Diagonal Loading for Norm Constrained Beamforming}

\name{Manan Mittal $^1$, Ryan M. Corey $^{2}$, John R. Buck $^3$, Andrew C. Singer $^1$ \thanks{Funded in part by US Navy, Office of Naval Research under award N00014-23-1-2133}}
\address{Stony Brook University$^1$, University of Illinois Chicago$^2$, University of Massachusetts Dartmouth$^3$}

\maketitle

\begin{abstract}
Reliable adaptive beamforming is critical for large microphone arrays operating in highly dynamic acoustic environments. In scenarios characterized by fast-moving talkers and interferers, the available sample support for estimating the spatial correlation matrix is often snapshot-deficient. This deficiency, coupled with array imperfections, degrades the White Noise Gain (WNG), leading to severe target signal cancellation. To ensure stable and robust beamforming, we propose a novel adaptive diagonal loading method that guarantees the WNG remains strictly within specified bounds. By leveraging the Kantorovich inequality, we map the desired WNG to a strict upper bound on the condition number of the correlation matrix. Furthermore, we present three estimation techniques for the adaptive loading level, ranging from trace-based bounding to exact eigenvalue decomposition, offering scalable computational complexities of $\mathcal{O}(M)$, $\mathcal{O}(M^2)$, and $\mathcal{O}(M^3)$. Our approach demonstrates highly stable beamforming under fast-changing interference.
\end{abstract}

\section{Introduction}
Adaptive beamforming techniques, such as the Minimum Power Distortionless Response (MPDR) and the Minimum Variance Distortionless Response (MVDR) \cite{capon}, are cornerstone algorithms in real time audio signal processing for noise reduction, dereverberation, and speech enhancement. These techniques achieve high spatial resolution by adapting their spatial filter weights to the second-order statistics of the received acoustic data. However, ensuring the robustness of these adaptive beamformers remains a significant challenge, particularly when deploying large microphone arrays in dynamic, real-world environments characterized by fast-moving interferers and talkers.

The fundamental vulnerability of optimal adaptive beamforming lies in its reliance on the sample Spatial Correlation Matrix (SCM) \cite{gannot2017consolidated}. To accurately track a fast-moving acoustic scene, the observation window used to estimate the SCM must be kept exceedingly short. When the number of available snapshots (frames) is less than or comparable to the number of microphone elements, the SCM becomes poorly conditioned or mathematically rank-deficient. Sample matrix inversion under this snapshot deficiency causes the spatial weights to become highly erratic. This phenomenon is exacerbated by inevitable array imperfections, such as sensor positioning errors, gain mismatches, and phase perturbations \cite{levin2013robust}. Consequently, the adaptive beamformer exhibits extreme sensitivity to spatially uncorrelated noise, which manifests as a dramatic collapse in the White Noise Gain (WNG) and severe cancellation of the target signal \cite{cox1987robust, li2006robust}. 

The classical remedy to mitigate SCM ill-conditioning and control the weight vector norm is Diagonal Loading (DL) \cite{van2002optimum, elnashar2006further, mestre2005diagonal}. By adding a scaled identity matrix to the SCM prior to inversion, DL artificially inflates the spatial noise floor, effectively bounding the maximum condition number of the matrix. While standard DL is ubiquitous, selecting the optimal loading parameter $\mu$ is historically an ad-hoc process. Applying a fixed $\mu$ that is too large over-penalizes the adaptive degrees of freedom, transforming the MPDR beamformer back into a suboptimal delay-and-sum beamformer that fails to null strong interferers. Conversely, a $\mu$ that is too small fails to stabilize the matrix during severe snapshot deficiency, leading to signal distortion. Literature has proposed several robust beamforming techniques over the years to improve the beamformers response to mismatch \cite{bell2002bayesian, feldman2002projection, shahbazpanahi2003robust, vorobyov2003robust, lam2006bayesian, itzhak2024robust, mabande2009robust}. In this paper, we propose a dynamic, closed-form adaptive diagonal loading method that deterministically guarantees the WNG stays within specified bounds. By exploiting the strict mathematical relationship between the array's WNG, the array gain, and the condition number of the SCM via the Kantorovich inequality, we derive an exact analytical requirement for the loading parameter at every frame. 

Because computing the exact eigenvalues of the SCM to determine the necessary DL at every time step is computationally expensive for large arrays, we introduce three progressive bounding techniques for estimating the loading level. These techniques utilize trace-based bounding, the Gershgorin circle theorem, and exact eigenvalue decomposition, offering scalable computational complexities from $\mathcal{O}(M)$ to $\mathcal{O}(M^3)$. 

\section{Signal Model}
We consider a room acoustic environment capturing a group of $J$ active sound sources (target talkers and interferers) using an array of $M$ microphones. Following the narrowband multiplicative assumption in the Short-Time Fourier Transform (STFT) domain, the signal received at the $m$-th microphone can be expressed as:
\begin{equation}
    y_m[i,k] = \sum_{j=1}^{J} h_{m,j}[k] s_j[i,k] + v_m[i,k]
\end{equation}
where $i$ is the time frame index, $k$ is the frequency bin index, $s_j[i,k]$ is the STFT of the $j$-th source signal, $h_{m,j}[k]$ is the Acoustic Transfer Function (ATF) from source $j$ to microphone $m$, and $v_m[i,k]$ represents spatially uncorrelated additive sensor noise.

For brevity, we omit the frequency index $k$ in the subsequent vector formulation. Vectorizing across the $M$ microphones, the array signal model at frame $i$ is written as:
\begin{equation}
    \mathbf{y}[i] = \mathbf{H}\mathbf{s}[i] + \mathbf{v}[i]
\end{equation}
where $\mathbf{y}[i] \in \mathbb{C}^{M \times 1}$, $\mathbf{s}[i] \in \mathbb{C}^{J \times 1}$, and $\mathbf{H} \in \mathbb{C}^{M \times J}$ is the matrix of stacked acoustic transfer functions. For a target source of interest, we define the Relative Transfer Function (RTF) or steering vector as $\mathbf{d} \in \mathbb{C}^{M \times 1}$, normalized such that $\mathbf{d}^H \mathbf{d} = M$.

The MPDR beamformer seeks a weight vector $\mathbf{w}[i] \in \mathbb{C}^{M \times 1}$ that minimizes the output power while maintaining a distortionless response in the target direction:
\begin{equation}
    \min_{\mathbf{w}} \mathbf{w}^H \mathbf{R}_y \mathbf{w} \quad \text{s.t.} \quad \mathbf{w}^H \mathbf{d} = 1
\end{equation}
where $\mathbf{R}_y = \mathbb{E}[\mathbf{y}\mathbf{y}^H]$ is the theoretical SCM. The well-known optimal solution is given by $\mathbf{w}_{opt} = \frac{\mathbf{R}_y^{-1} \mathbf{d}}{\mathbf{d}^H \mathbf{R}_y^{-1} \mathbf{d}}$. 

In practice, the true SCM is unknown and must be approximated via a short sliding window to track moving sources: 
\begin{equation}
    \hat{\mathbf{R}}_y[i] = \frac{1}{L} \sum_{l=0}^{L-1} \mathbf{y}[i-l]\mathbf{y}^H[i-l]
\end{equation}
When the window length $L$ is small ($L < M$), $\hat{\mathbf{R}}_y[i]$ is rank-deficient. Its minimum eigenvalues approach zero, causing the condition number to approach infinity and the inverse $\hat{\mathbf{R}}_y^{-1}$ to heavily amplify minor estimation errors and uncorrelated noise.

\section{Proposed Method}

\subsection{WNG Bounds via Kantorovich Limits}
The robustness of a beamformer to uncorrelated noise is quantified by its White Noise Gain (WNG), defined as the ratio of the output SNR to the input SNR in a spatially white noise field,
\begin{equation}
W = \frac{\lvert\mathbf{w}^H \mathbf{d}\rvert}{\lvert\mathbf{w}^H \mathbf{w}\rvert} = \frac{1}{\lvert\mathbf{w}^H \mathbf{w}\rvert}
\end{equation}

where the second equality holds due to the distortionless constraint $\mathbf{w}^H \mathbf{d} = 1$. In a snapshot-deficient MPDR beamformer, the norm of the weight vector $\|\mathbf{w}\|^2$ spikes dramatically, causing $W$ to plummet. To guarantee stable beamforming, we must enforce a strict lower bound, $W \geq W_{\min}$.

The weight norm for the MPDR beamformer can be rewritten in terms of the SCM:
\begin{equation}
    \mathbf{w}^H \mathbf{w} = \frac{\mathbf{d}^H \mathbf{R}_y^{-2} \mathbf{d}}{(\mathbf{d}^H \mathbf{R}_y^{-1} \mathbf{d})^2}
\end{equation}
To strictly bound this ratio, we leverage the Kantorovich inequality \cite{kantorovich1948functional}. For any Hermitian positive-definite matrix $\mathbf{R}$ with condition number $\kappa = \lambda_{\max} / \lambda_{\min}$, and for any non-zero vector $\mathbf{x}$, the general inequality states:
\begin{equation}
    \frac{(\mathbf{x}^H \mathbf{x})^2}{(\mathbf{x}^H \mathbf{R} \mathbf{x})(\mathbf{x}^H \mathbf{R}^{-1} \mathbf{x})} \geq \frac{4\kappa}{(\kappa+1)^2}
\end{equation}
To apply this to our beamforming problem, let $\mathbf{R} = \mathbf{R}_y$ and define the vector $\mathbf{x} = \mathbf{R}_y^{-1/2} \mathbf{d}$. Substituting these into the component terms of the inequality yields:
\begin{equation}
    \mathbf{x}^H \mathbf{x} = \mathbf{d}^H \mathbf{R}_y^{-1} \mathbf{d}
\end{equation}
\begin{equation}
    \mathbf{x}^H \mathbf{R}_y \mathbf{x} = \mathbf{d}^H \mathbf{R}_y^{-1/2} \mathbf{R}_y \mathbf{R}_y^{-1/2} \mathbf{d} = \mathbf{d}^H \mathbf{d} = M
\end{equation}
\begin{equation}
    \mathbf{x}^H \mathbf{R}_y^{-1} \mathbf{x} = \mathbf{d}^H \mathbf{R}_y^{-2} \mathbf{d}
\end{equation}
Plugging these expanded terms back into the Kantorovich inequality gives:
\begin{equation}
    \frac{(\mathbf{d}^H \mathbf{R}_y^{-1} \mathbf{d})^2}{M (\mathbf{d}^H \mathbf{R}_y^{-2} \mathbf{d})} \geq \frac{4\kappa}{(\kappa+1)^2}
\end{equation}
Recognizing from our earlier definitions that $W = \frac{(\mathbf{d}^H \mathbf{R}_y^{-1} \mathbf{d})^2}{\mathbf{d}^H \mathbf{R}_y^{-2} \mathbf{d}}$, we obtain the relationship:
\begin{equation}
    \frac{W}{M} \geq \frac{4\kappa}{(\kappa+1)^2}
\end{equation}
Let $A_G = M / W_{\min}$ represent the strict array gain limit (i.e., the maximum allowable degradation relative to the optimal delay-and-sum Array Gain). To guarantee $W \geq W_{\min}$, we set $M / W = A_G$ and solve the inequality for the maximum allowable condition number $\kappa_{\max}$:
\begin{equation}
    \kappa_{\max} = (2A_G - 1) + 2\sqrt{A_G(A_G - 1)}
\end{equation}
This is a powerful deterministic result: by actively limiting the condition number of the estimated SCM, we implicitly and strictly control the WNG of the resulting adaptive beamformer.

\subsection{Adaptive Diagonal Loading Estimation}
To actively constrain the SCM's condition number to $\kappa_{\max}$, we apply a dynamic diagonal loading factor $\mu[i]$ at every frame:
\begin{equation}
    \mathbf{Q}[i] = \hat{\mathbf{R}}_y[i] + \mu[i] \mathbf{I}
\end{equation}
Let $\lambda_{\max}$ and $\lambda_{\min}$ be the maximum and minimum eigenvalues of the unloaded sample matrix $\hat{\mathbf{R}}_y[i]$. The eigenvalues of the loaded matrix $\mathbf{Q}[i]$ are shifted by $\mu[i]$, yielding a new condition number:
\begin{equation}
    \kappa_{loaded} = \frac{\lambda_{\max} + \mu[i]}{\lambda_{\min} + \mu[i]}
\end{equation}
To satisfy $\kappa_{loaded} \leq \kappa_{\max}$, we solve for the exact required loading multiplier:
\begin{equation}
    \mu[i] = \max \left(0, \frac{\lambda_{\max} - \kappa_{\max}\lambda_{\min}}{\kappa_{\max} - 1} \right)
\end{equation}
This formulation ensures that we apply the absolute minimum amount of diagonal loading necessary to preserve the requested WNG, thereby preserving the beamformer's ability to place deep nulls on interferers. 

\subsection{Complexity Scalable Estimation Modes}
Calculating $\mu[i]$ requires knowledge of the SCM's extreme eigenvalues at every frame. Because Exact Eigenvalue Decomposition (EVD) is computationally prohibitive ($\mathcal{O}(M^3)$) for arrays with many elements operating at high sampling rates, we propose three scalable estimation techniques:

\begin{enumerate}
    \item \textbf{Trace Mode ($\mathcal{O}(M)$):} The sum of the eigenvalues equals the trace of the matrix. Since SCMs are positive semi-definite, we can formulate a rapid, strictly conservative upper bound: $\lambda_{\max} \leq \text{Tr}(\hat{\mathbf{R}}_y)$. We assume worst-case snapshot deficiency where $\lambda_{\min} \approx 0$. This mode is extremely fast but results in slightly heavier diagonal loading than strictly necessary.
    
    \item \textbf{Gershgorin Mode ($\mathcal{O}(M^2)$):} This mode utilizes the Gershgorin Circle Theorem to place tighter bounds on the eigenspectrum without performing a full decomposition. Every eigenvalue of $\hat{\mathbf{R}}_y$ lies within at least one Gershgorin disc $D( \hat{R}_{m,m}, R_m)$, where the radius is the sum of the absolute off-diagonal elements in that row: $R_m = \sum_{j \neq m} \lvert \hat{R}_{m,j}\rvert$. We estimate the bounds as:
    \begin{align}
        \lambda_{\max} &\leq \max_m \left( \hat{R}_{m,m} + R_m \right) \\
        \lambda_{\min} &\geq \max \left( 0, \min_m \left( \hat{R}_{m,m} - R_m \right) \right)
    \end{align}
    This provides an excellent trade-off, offering tighter loading limits at moderate complexity.
    
    \item \textbf{Exact EVD ($\mathcal{O}(M^3)$):} For smaller arrays or systems with high computational budgets, exact eigenvalues are extracted. This guarantees the theoretically optimal $\mu[i]$, providing the highest possible interference suppression while exactly adhering to the WNG limit.
\end{enumerate}

\subsection{WNG Bounds in the GSC Framework}
In the direct MPDR formulation, the target distortionless constraint and the adaptive degrees of freedom are entangled within the same weight vector. Alternatively, the Generalized Sidelobe Canceller (GSC) architecture orthogonalizes these components. The overall weight vector in the GSC is defined as:
\begin{equation}
    \mathbf{w}_{gsc} = \mathbf{w}_q - \mathbf{B} \mathbf{w}_a
\end{equation}
where $\mathbf{w}_q = \mathbf{d}/M$ is the fixed quiescent weight vector satisfying the target constraint, $\mathbf{B} \in \mathbb{C}^{M \times (M-1)}$ is the blocking matrix such that $\mathbf{B}^H \mathbf{d} = \mathbf{0}$ and $\mathbf{B}^H \mathbf{B} = \mathbf{I}$, and $\mathbf{w}_a \in \mathbb{C}^{(M-1) \times 1}$ is the adaptive noise cancellation weight vector.

The adaptive noise cancellation weight vector is traditionally computed as $\mathbf{w}_a = \mathbf{R}_n^{-1} \mathbf{r}_{qn}$, where $\mathbf{R}_n = \mathbf{B}^H \hat{\mathbf{R}}_y \mathbf{B}$ is the noise correlation matrix and $\mathbf{r}_{qn} = \mathbf{B}^H \hat{\mathbf{R}}_y \mathbf{w}_q$ is the cross-correlation vector. Let $p_q = \mathbf{w}_q^H \hat{\mathbf{R}}_y \mathbf{w}_q$ denote the quiescent output power tracked over the sliding window.

To enforce the Kantorovich-derived WNG bounds without explicitly reconstructing the full spatial correlation matrix $\hat{\mathbf{R}}_y$, we define a unitary transformation matrix $\mathbf{T} = [\sqrt{M}\mathbf{w}_q, \mathbf{B}]$. Because $\mathbf{T}^H \mathbf{T} = \mathbf{I}$, the transformed matrix $\tilde{\mathbf{R}} = \mathbf{T}^H \hat{\mathbf{R}}_y \mathbf{T}$ inherently shares the exact same eigenvalues as $\hat{\mathbf{R}}_y$. By leveraging the orthogonal properties of the GSC, we can construct $\tilde{\mathbf{R}}$ directly from the continuously tracked components:
\begin{equation}
    \tilde{\mathbf{R}} = \begin{bmatrix} M p_q & \sqrt{M} \mathbf{r}_{qn}^H \\ \sqrt{M} \mathbf{r}_{qn} & \mathbf{R}_n \end{bmatrix}
\end{equation}

Because the eigenspectrum is perfectly preserved, the extreme eigenvalues $\lambda_{\max}$ and $\lambda_{\min}$ can be estimated from $\tilde{\mathbf{R}}$ using the previously defined scalable modes (Trace, Gershgorin, or Exact EVD). Consequently, the requisite adaptive diagonal loading factor $\mu[i]$ is identical to that of the direct MPDR formulation for the Trace and EVD modes. Note, the Gershgorin estimates depend on the choice of basis functions, and will generally result in different diagonal loading estimates. This is shown in the simulations. 

The WNG-constrained beamformer is then realized by applying this condition-bounding load solely to the noise correlation matrix prior to inversion:
\begin{equation}
    \mathbf{w}_a = (\mathbf{R}_n + \mu[i] \mathbf{I})^{-1} \mathbf{r}_{qn}
\end{equation}

This formulation demonstrates that the proposed dynamic loading technique is structurally agnostic, providing the exact same deterministic WNG guarantees and stability whether applied directly to the sample matrix inversion or within the partitioned GSC framework.

\section{Simulations}

\subsection{Simulation Setup}
We evaluate the proposed adaptive diagonal loading strategies using a simulated uniform linear array (ULA) consisting of $M=15$ microphones with half-wavelength spacing at a center frequency of $f_0 = 1000$ Hz. To rigorously test the tracking capabilities and robustness of the algorithms, we simulate a highly dynamic ``birth-death'' spatial interference scenario over $T=20000$ snapshots. In this scenario, up to two statistically independent interferers randomly appear, remain active for a duration, and disappear. 

To prevent trivial interference scenarios or impossible target separation, the interferers are strictly confined to an angular grid where the target's normalized quiescent beampattern response falls between $-13$ dB and $-3$ dB. This may be typical in cocktail party scenario where multiple closely spaced talkers may need to be separated. The dynamic interferers are generated with an Interference-to-Noise Ratio (INR) of $7$ dB. The target signal is fixed at broadside ($90^\circ$) with a Signal-to-Noise Ratio (SNR) of $-5$ dB. To induce severe snapshot deficiency, the sample Spatial Correlation Matrix (SCM) is tracked using a sliding rectangular window of $L = 37$ snapshots ($L \approx 2.5M$). For an array of $M=15$, the maximum theoretical WNG is $10 \log_{10}(15) \approx 11.76$ dB. To allow for adaptive interference nulling while preventing target cancellation, we define a strict WNG lower bound of $W_{\min} = 10 \log_{10}(M) - 3 \approx 8.76$ dB.

We compare the three proposed complexity modes—Trace, Gershgorin, and Exact Eigenvalue Decomposition (EVD)—against the classical post-hoc weight scaling method proposed by Cox \textit{et al.} \cite{cox1987robust}, and an Omniscient Capon beamformer. The Omniscient Capon utilizes the exact, theoretical underlying ECM at every snapshot and serves as the absolute upper bound for achievable performance.
\subsection{Results and Discussion}

\begin{figure}[t]
    \centering
    \includegraphics[width=\columnwidth]{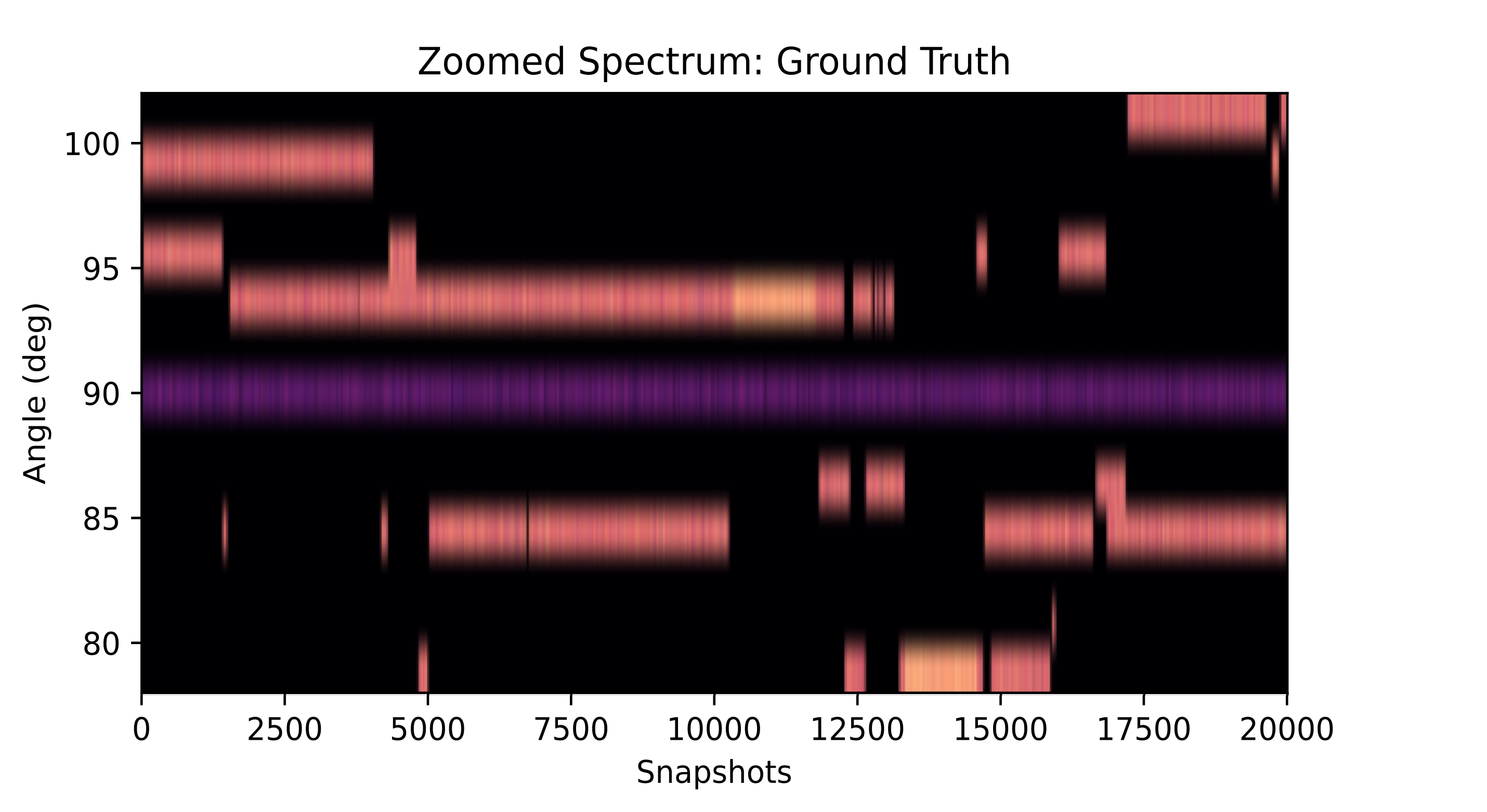} % Update path
    \caption{Ground Truth Spatial spectrum over time for a single trial, demonstrating the dynamic birth-death interferers and the broadside target.}
    \label{fig:spatial_spectrum}
\end{figure}

The ground truth scanned spatial response over a single trial is shown in \ref{fig:spatial_spectrum}. The beamformers successfully place and dynamically update deep nulls as the birth-death interferers transition, without suppressing the broadside target.

\begin{figure}[t]
    \centering
    \includegraphics[width=\columnwidth]{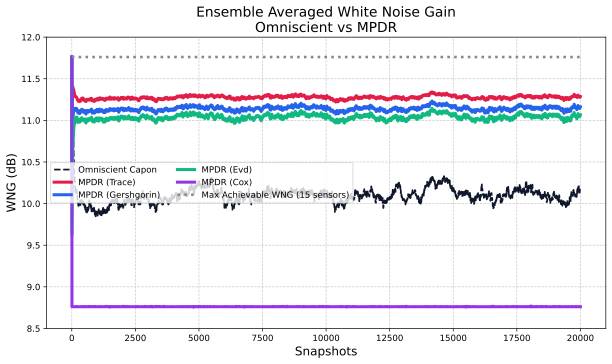} % Update path
    \caption{Ensemble White Noise Gain (WNG). All proposed pre-inversion conditioning methods actively bound the WNG above the specified $8.76$ dB limit.}
    \label{fig:wng_ensemble}
\end{figure}

The primary objective of the proposed Kantorovich-bounded loading is to guarantee WNG stability. Fig. \ref{fig:wng_ensemble} plots the ensemble WNG over time. Under tight snapshot deficiency, standard sample matrix inversion causes the weight vector norm to explode, resulting in dramatic target cancellation. As designed, the Trace, Gershgorin, and EVD modes strictly and actively constrain the WNG above the $8.76$ dB threshold at every frame. 

\begin{figure}[t]
    \centering
    \includegraphics[width=\columnwidth]{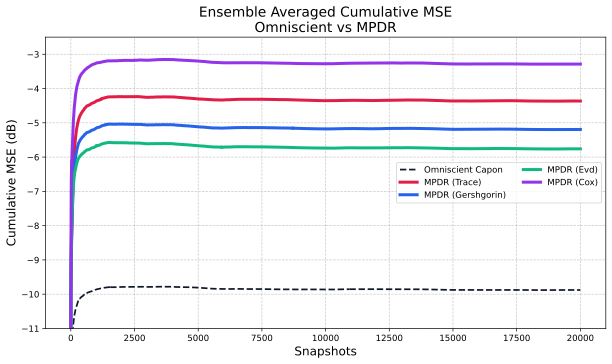} % Update path
    \caption{Ensemble cumulative Mean Squared Error (MSE) for the MPDR formulation.}
    \label{fig:mse_ensemble}
\end{figure}

\begin{figure}[t]
    \centering
    \includegraphics[width=\columnwidth]{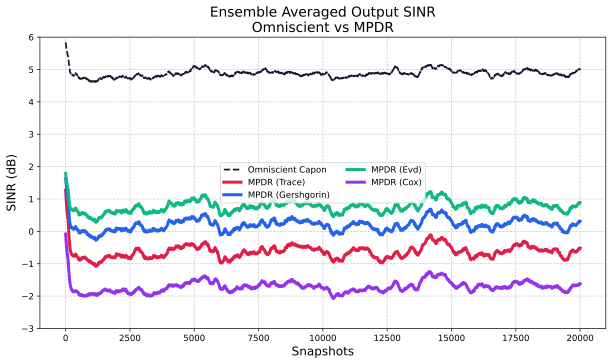} % Update path
    \caption{Ensemble output SINR. The Exact EVD and Gershgorin modes track closely to the Omniscient baseline.}
    \label{fig:sinr_ensemble}
\end{figure}

The ensemble cumulative Mean Squared Error (MSE) and output Signal-to-Interference-plus-Noise Ratio (SINR) are presented in Fig. \ref{fig:mse_ensemble} and Fig. \ref{fig:sinr_ensemble}, respectively. Among the realizable methods, a clear performance hierarchy emerges: Exact EVD $>$ Gershgorin $>$ Trace $>$ Cox. 

The \textbf{Exact EVD} mode achieves the highest output SINR. By extracting the true extreme eigenvalues, it applies the exact minimal diagonal loading necessary, preserving the maximum possible degrees of freedom for deep interference nulling. The \textbf{Gershgorin} mode performs nearly identically to the EVD mode, successfully suppressing the dynamic interferers while drastically reducing the computational burden to $\mathcal{O}(M^2)$. The \textbf{Trace} mode ($\mathcal{O}(M)$) serves as a rapid, strictly conservative bound; because it slightly overestimates the required loading, it behaves closer to a delay-and-sum beamformer, resulting in a marginally lower SINR but absolute WNG stability. 

In contrast, while the Cox method attempts to restore WNG via post-hoc scaling of the weight vector's null-space projection, this ad-hoc geometric adjustment disrupts the optimality of the spatial filter. This results in significantly worse cumulative MSE and slower convergence during interferer transitions compared to our proposed pre-inversion conditioning.

\subsection{Architecture Equivalence: MPDR vs. GSC}

We also evaluated the algorithms within the Generalized Sidelobe Canceller (GSC) architecture. Mathematically, the EVD, Trace, and Cox modes are perfectly invariant under the unitary transformation mapping the direct MPDR to the GSC framework. Therefore, they yield identical beamforming weights and performance in both architectures. 

\begin{figure}[t]
    \centering
    \includegraphics[width=\columnwidth]{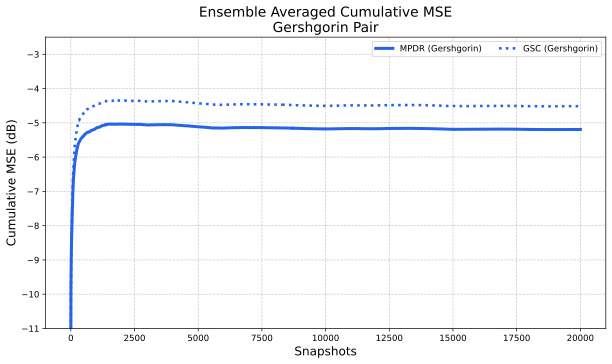} % Update path
    \caption{Performance comparison of the Gershgorin mode between the direct MPDR and GSC architectures, highlighting the basis-dependent nature of the eigenvalue bounds.}
    \label{fig:gershgorin_comparison}
\end{figure}

However, as illustrated in Fig. \ref{fig:gershgorin_comparison}, the \textbf{Gershgorin} mode exhibits divergent behavior between the MPDR and GSC formulations. This discrepancy arises because the Gershgorin circle bounds are inherently basis-dependent. The unitary blocking matrix $\mathbf{B}$ applied in the GSC alters the distribution of matrix energy between the diagonal and off-diagonal elements of the partitioned correlation matrix. Because the Gershgorin radii are defined by the sum of the absolute off-diagonal elements, this transformation tightens or loosens the estimated eigenvalue bounds depending on the instantaneous snapshot data, resulting in slight variations in the applied loading parameter $\mu$ compared to the direct MPDR domain.

\section{Conclusion}
In this paper, we proposed a novel WNG-constrained adaptive diagonal loading approach tailored for snapshot-deficient scenarios and highly dynamic acoustic environments. By establishing an analytic bound using the Kantorovich inequality, our method actively and deterministically constrains the condition number of the spatial correlation matrix. This mathematically guarantees that the beamformer's White Noise Gain remains strictly within specified bounds, preventing the severe target signal cancellation commonly observed in standard MPDR and MVDR beamformers. Furthermore, we introduced three progressive eigenvalue bound estimation techniques—Trace, Gershgorin, and Exact EVD—that provide flexible trade-offs between computational complexity and strict WNG adherence. Our simulations demonstrate that this approach yields highly stable, robust beamforming that outperforms classical post-hoc scaling methods, making it highly suitable for large microphone arrays operating in real-world, low-latency audio applications. It also provides a principled approach to diagonal loading for neural estimated covariance matrices which have become a standard estimation technique. 

\bibliographystyle{IEEEbib}
\bibliography{refs}

\end{document}